# Revisiting the predictability of the Haicheng and Tangshan earthquakes

Didier Sornette[1-3], Euan Mearns[1] and Spencer Wheatley[1]

[1] ETH Zurich, Chair of Entrepreneurial Risks, D-MTEC, Scheuchzerstrasse 7, CH-8092 Zurich, Switzerland
[2] Institute of Risk Analysis, Prediction and Management, Academy for Advanced Interdisciplinary Studies, Southern University of Science and Technology, Shenzhen, 518055, China
[3] Tokyo Tech World Research Hub Initiative (WRHI), Institute of Innovative Research Tokyo Institute of Technology, Tokyo, Japan

**Abstract**: We analyse the compiled set of precursory data that were reported to be available in real time before the $M_S7.5$ Haicheng earthquake in Feb. 1975 and the $M_S7.6$-7.8 Tangshan earthquake in July 1976. We propose a robust and simple coarse-graining method consisting in aggregating and counting how all the anomalies together (geodesy, levelling, geomagnetism, soil resistivity, Earth currents, gravity, Earth stress, well water radon, well water level) develop as a function of time. We demonstrate a strong evidence for the existence of an acceleration of the number of anomalies leading up to the major Haicheng and Tangshan earthquakes. In particular for the Tangshan earthquake, the frequency of occurrence of anomalies is found to be well described by the log-periodic power law singularity (LPPLS) model, previously proposed for the prediction of engineering failures and later adapted to the prediction of financial crashes. Based on a mock real-time prediction experiment, and simulation study, we show the potential for an early warning system with lead-time of a few days, based on this methodology of monitoring accelerated rates of anomalies.

## 1-Introduction

In the period 1966 to 1976, the Chinese engaged in the biggest earthquake monitoring program ever conceived. It began with the Xingtai earthquake sequence, of which the two most powerful occurred on (UTC) 7th March ($M_W$=6.5) and 22nd March 1966 ($M_W$=6.8). The program more or less ended with the sequence of Tangshan earthquakes that began on 28 July 1976 and ended on 27 November 1977 after 30 M>5 shocks had occurred (Jiang and Chen, 1982, Table 2.6). Following Xingtai, Chinese premier Enlai Zhou was concerned by the scale of casualties and instructed Siguang Li, China's most senior geologist, to establish a monitoring program with the aim of predicting earthquakes and offering some advance warning to the people.

Stories of strange phenomena preceding earthquakes were engrained in Chinese folklore, such as strange behaviour of animals, turbid water in wells and strange fogs and clouds. The idea was to combine a range of observations and simple measurements made by amateur groups and to combine these with more sophisticated measurements made by professional groups. In this way, China could capitalise on one of its main resources, i.e. people, but this would also engage the population in the prediction program giving the people heightened awareness of earthquake risks.

At its peak, the program engaged 35,000 amateur groups located in schools, factories and public buildings and consequently a large and unique set of physical measurements of anomalies and human observations were acquired over the decade. Some of this data is published in graph form in post-mortem reports published after the Haicheng, Tangshan and Songpan earthquakes (Zhu and Wu, 1982; Jiang and Chen, 1982; Sichuan Province earthquake office, 1979), but accessing the raw data that lies under the charts is not straightforward. Amongst other things, the Chinese language presents a barrier.

In 1981, Jian-zhong Zheng, a Chinese scientist previously affiliated with the Tokyo Institute of Technology (employed at the time of publication at the Institute of Geophysics, Chinese Academy of Sciences, Beijing), published a paper that described some of the anomalies recorded prior to the Haicheng (February 1975) and Tangshan (July 1976) earthquakes (Zheng, 1981). Published in Japanese, the paper has an English abstract and two English language appendices documenting anomalies from the Haicheng and Tangshan earthquakes respectively. We saw here an opportunity to perform a mathematical analysis of the precursor data using the Finite Time Singularity model (Anifrani et al., 1995; Gluzman and Sornette, 2002; Ide and Sornette, 2002; Sornette, 2002; Sornette and Sammis, 2002). Our aim is to see if we could have predicted these earthquakes based on acceleration in anomalies leading up to the climax.

As the next section illustrates, the precursors that were monitored in real time are very diverse, and exhibit strong heterogeneity in space and time. The large variability of their spatial distributions, times of occurrence and amplitudes, confused the Chinese seismologists in charge of the earthquake monitoring program. The guiding idea of the present study is to see if coarse-graining this wealth of noisy information might have brought some meaning to their development suggesting that, together, they might have been sufficient to provide useful predictions for decision makers to act upon. In other words, the key idea is to avoid being swamped by the details and the corresponding overwhelming variability and try to extract a robust signal. This strategy has been successful for the prediction of failures of engineering structures, also characterized by enormous variability of the recorded signals (Anifrani et al., 1995; Sornette, 2002). The coarse-graining strategy gets its inspiration from the Renormalisation Group theory of critical phenomena (Goldenfeld, 1992; Sornette, 2004).

Our approach is reminiscent of the accelerated moment release method (Bowman et al., 1998; Mignan, 2011; Guilhem et al. 2013), based on the hypothesized increase in seismicity prior to a major earthquake. We stress however the fundamental difference in the fact that the accelerated moment release method only uses seismicity while our study is built on eight non-seismic variables (geodesy, levelling, geomagnetism, soil resistivity, Earth currents, gravity, Earth stress, well water radon, well water level). The present work is thus more in the spirit of Johansen et al. (1996; 2000), but with a much larger set of physical variables to aggregate upon.

Section 2 describes important background information on the different types as well as on the spatio-temporal organization of the hypothesized precursors to the Haichang and Tangshan earthquakes. Section 3 presents our analysis of the accelerated aggregate rate of these precursors before the Haicheng and Tangshan earthquakes, including an ex-post forecasting exercise as well as statistical tests of the predictability of Tangshan earthquake. Section 4 concludes by stressing the role of quantitative data aggregation and skilled display of technical information for diagnostic impending ruptures.

## 2-Background of precursors for the Haicheng and Tangshan earthquakes

For Haicheng, six anomaly classes are reported and numbers in brackets give the number of reported anomalies:
1) geomagnetism    (2)
2) resistivity    (2)
3) Earth currents    (13)
4) gravity    (2)
5) Earth stress    (8)
6) radon    (16)

Of these, only three have a reasonable quantity of data, namely Earth currents, Earth stress and radon. We do not really have sufficient information from Haicheng to conduct meaningful analysis and we shall therefore focus on Tangshan where more data is available:

1) geodimeter    (2)
2) levelling    (11)
3) geomagnetism    (11)
4) resistivity    (20)
5) Earth currents    (15)

6) gravity                    (14)
7) Earth stress               (31)
8) radon                      (29)
9) well water level           (21)

Zheng (1981) reports precursor time in days, epicentral distance in kilometres and the name of the observation station. This has enabled us to compile maps of anomaly distributions that we shall get to shortly. First, we will clarify what some of these anomalies mean.

Resistivity is measured between four steel electrodes hammered into the ground. An AC current is applied and the soil resistivity is computed.

Earth currents, also known as telluric currents, record spontaneous electric currents in the ground. Two steel electrodes are hammered into the ground, joined by a wire, and an ammeter records the current, if any.

Earth stress is measured by a special transducer planted in the soil and records changes in the stress condition of the soil.

Radon in China was measured in well water and not air. $^{222}$Rn concentrations are measured based on radioactivity using a scintillation counter. This is a relatively simple hand-held device enabling measurements to be made in amateur monitoring posts. The simplicity is derived from the fact that Rn is a gas that is separated from the water prior to analysis.

Most of the anomaly classes listed above may be viewed as trending anomalies, i.e. the value measured may vary with time. Ideally, these should be set against baseline values where a baseline is established prior to earthquake conditions and then something happens to cause a change in value from that baseline. Once established, then observations may be made about the change in trend, e.g. jumps, spikes and turning points. Zheng (1981) sometimes reports more than one anomaly per station, for example Changli has four resistivity anomalies. Zheng did not publish graphs that would allow assessment of the judgements applied.

Qian et al., (1997) published resistivity logs from the Tangshan area and this offers the opportunity to scrutinise Zheng's recordings against reality (Figure 1). What we find is that each locality has a relatively flat baseline three years prior to the event but then we begin to see baseline drift towards lower values. In each case, Zheng's recordings do mark the beginning of baseline drift. Subsequent recordings sometimes mark spikes or turning points but it is not always obvious what observations are being recorded.

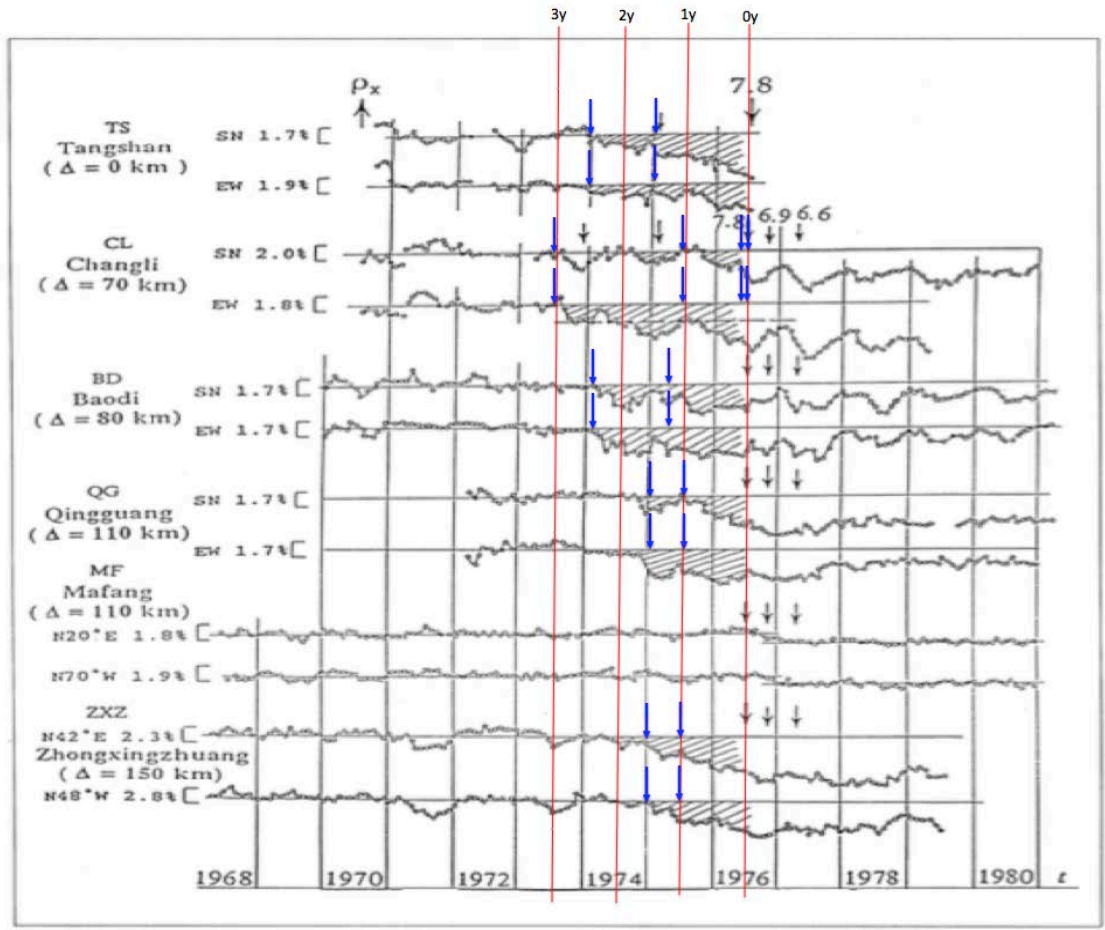

Figure 1. Resistivity logs from 6 stations in the Tangshan area as reported by Qian et al. (1997, their figure 3). Each station has N-S and E-W orientated devices. Zheng (1981) reports data for five of these stations. He does not report data for Mafang because there is no anomaly at that station. The blue arrows mark the anomaly times reported by Zheng (1981) in each case, Zheng's data marks the first appearance of a trend on the logs.

One important observation to make from Figure 1 is clear seasonal variation in some of the localities, for example Changli and Baodi. This is perhaps not surprising since resistivity may respond to rainfall (soil moisture), temperature and frozen ground. This caused us to temporarily doubt that resistivity, well water level and radon concentrations were responding to earthquake conditions at all. All may have been responding to climatic conditions and an uncommonly dry period did precede the Tangshan earthquake that would cause well water to fall and radon concentrations to rise. However, when we plot the data on a map (see below and Mearns and Sornette, 2020), we find that these anomalies also align with structural elements and so conclude that earthquake signals may be superimposed upon other effects such as seasonal weather variations.

One further question is to what extent the data of Zheng (1981) is complete. This was discussed in his paper where he thought that it should be relatively so. We can confirm this is not the case. Zheng reports Earth current data from 13 measuring stations and presumably only from stations that registered an anomaly. In the SSB (State Seismological Bureau) post-mortem report, Jiang and Chen (1982) publish a map showing Earth current anomaly stations and distinguish between those with an anomaly and those without. This map shows 41 measuring sites, 26 with anomalies and 15 without. This is roughly twice as many sites with anomalies as reported by Zheng (1981). However, the distributions of the anomalies from the two sources are very similar. Zheng's localities appear on the SSB map.

We now proceed to examine the temporal and spatial distribution of the anomalies. Figure 2 shows the time of first appearance of the anomalies on the y-axis and the time distribution of the anomalies as listed by Zheng (1981) along the x-axis. Panel (a) plots all the data for Tangshan (Zheng 1981 appendix 2) while panel (b) plots a subset (see below). The order in the legend records the sequential order of first appearance with ground water anomalies appearing first and Earth currents appearing last.

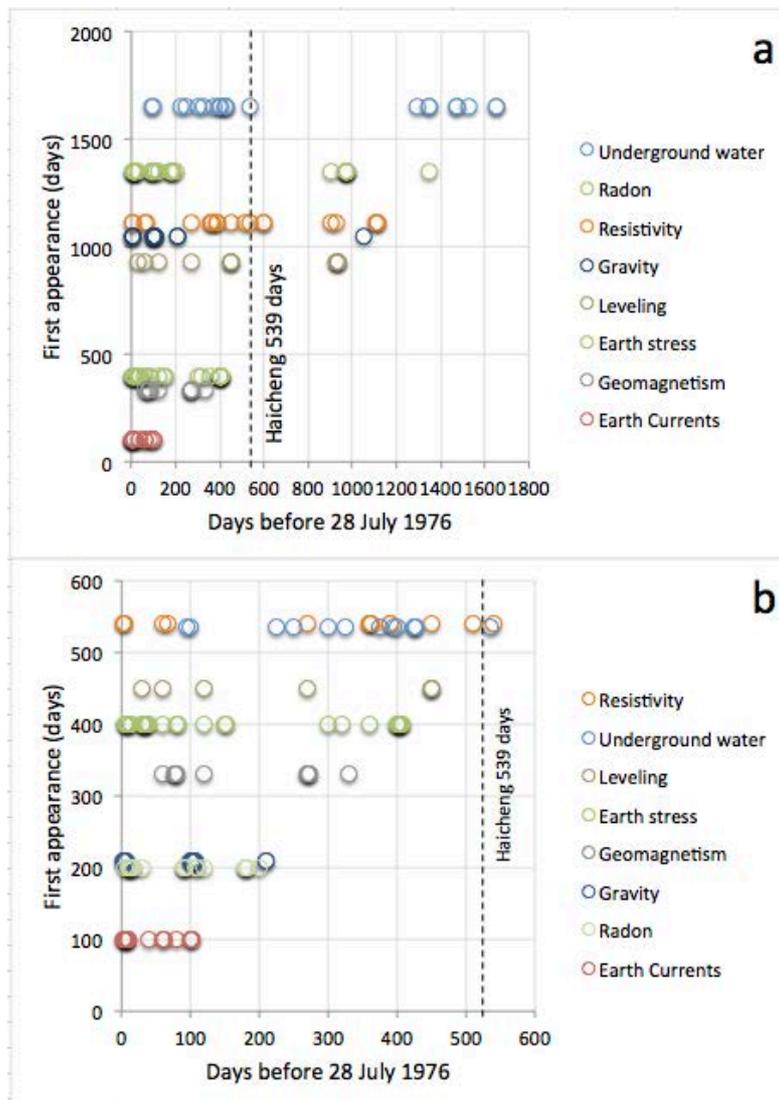

Figure 2. The distributions of anomaly times plotted along the x-axis versus the time of first appearance plotted on the y-axis. a) plots all the data and b) plots the sub-set of data with a time filter set at 540 days. This marks the reappearance of anomalies in Tangshan after the Haicheng earthquake. Note that the order of first appearnce in (b) is different to (a) since the former is adjusted to reflect the order seen in the post 540 day subset.

One notable feature from Figure 2a is the bi-modal distribution of water, radon, resistivity, gravity and levelling anomalies. There is an earlier group that disappears at ~ 900 days to re-appear again at ~ 539 days, coincident with the time of the Haicheng earthquake. Chinese seismologists have told us that, following the Bohai earthquake of 1969, stress indicators appeared in the vicinity of both Yingkou (close to Haicheng) and Tangshan. This was called the "aftereffect anomaly field" (Chengmin Wang, personal communication), a term used to represent the various fields (such as stress, strain, fluids, electric currents, and so on) thought to diffuse and potentially influence or signal the nucleation of future earthquakes as a result of a previous earthquake. What we may be seeing in the anomaly distribution is the reduction in this "aftereffect anomaly field" in Tangshan as it became focused on Haicheng. And then after the Haicheng earthquake, the aftereffect anomaly field re-appeared in Tangshan.

For our analysis and on the maps below, we have set a filter at 540 days so that we only use anomaly data that is directly linked to the gestation of the Tangshan earthquake. Setting the filter at 540 days, we have re-ordered the data in Figure 2b that gives a new order for the first appearance of anomalies.

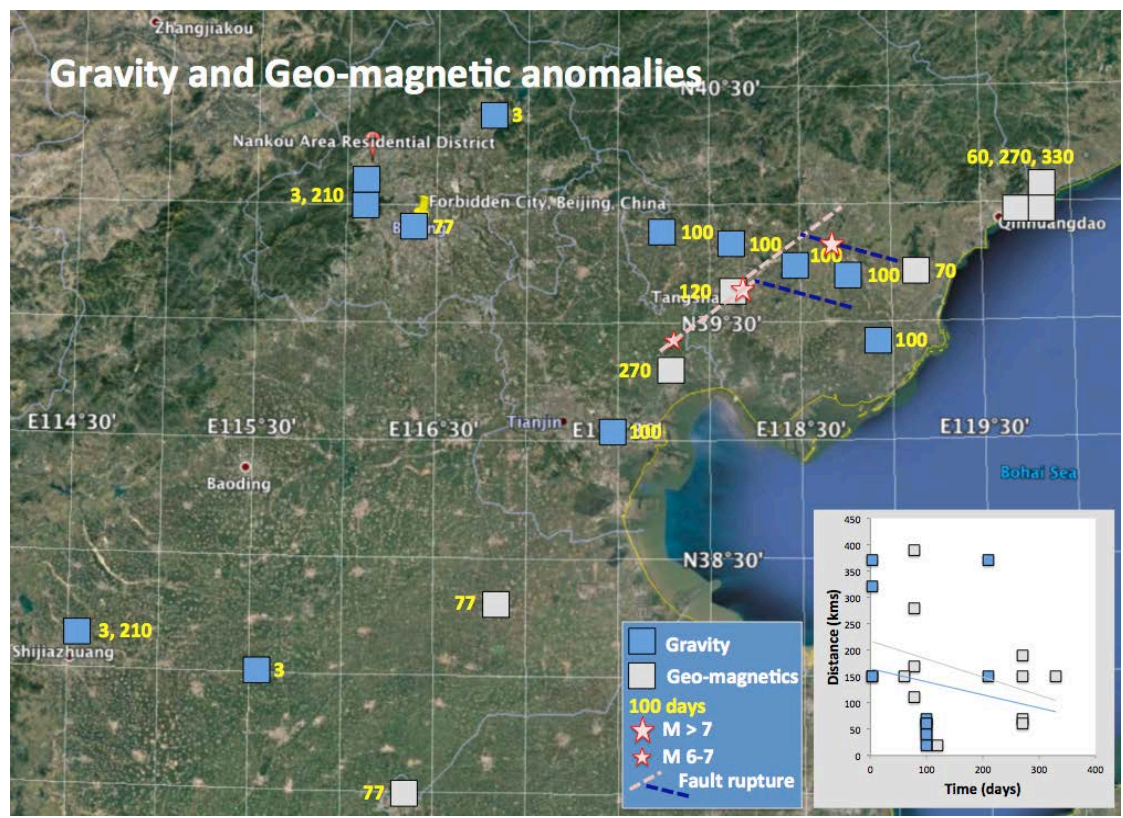

Figure 3. Map showing the spatial distribution of gravity and geomagnetic anomalies. The yellow numbers give the anomaly time in days before the main shock of the Tangshan earthquake (28 July 1976). The stars show the locations of the three main shocks, Tangshan (28 July 1976) in the middle, Luanxian to the NW (also on 28 July 1976) and Ninghe to the SE (15 November 1976). The dashed lines show our interpretation of the Tangshan transfer fault system where pink denotes a dextral strike-slip leg and dark blue denotes a normal fault leg. Base image from Google Earth. The dappled beige colour reflects the flat surface of the densely populated Bohai Bay rift basin. The dark green reflects the mountainous, forested and dissected Yanshan Range. The boundary between the two marks the boundary of the Bohai Bay basin. The inset graph plots the distance as a function of time of the anomalies. The straight lines suggest a weak inverse correlation between time and distance.

Figure 3 shows the distribution of gravity and geomagnetic anomalies. There is a slight concentration of anomalies around Tangshan that may follow the fault lines as we reconstruct them (Mearns and Sornette, 2020). Equally, theses anomalies are spread all over the map. There is a tendency for these anomalies to follow the basin margin and for the geomagnetic anomalies to be distributed towards the east. The inset graph shows both of these anomaly classes have a weak inverse correlation between time and distance. In other words, as the earthquake approached, the anomalies became more distant. This observation confused the Chinese seismologists at the time and contributed to the failure to predict the Tangshan earthquake (Mei, 1986).

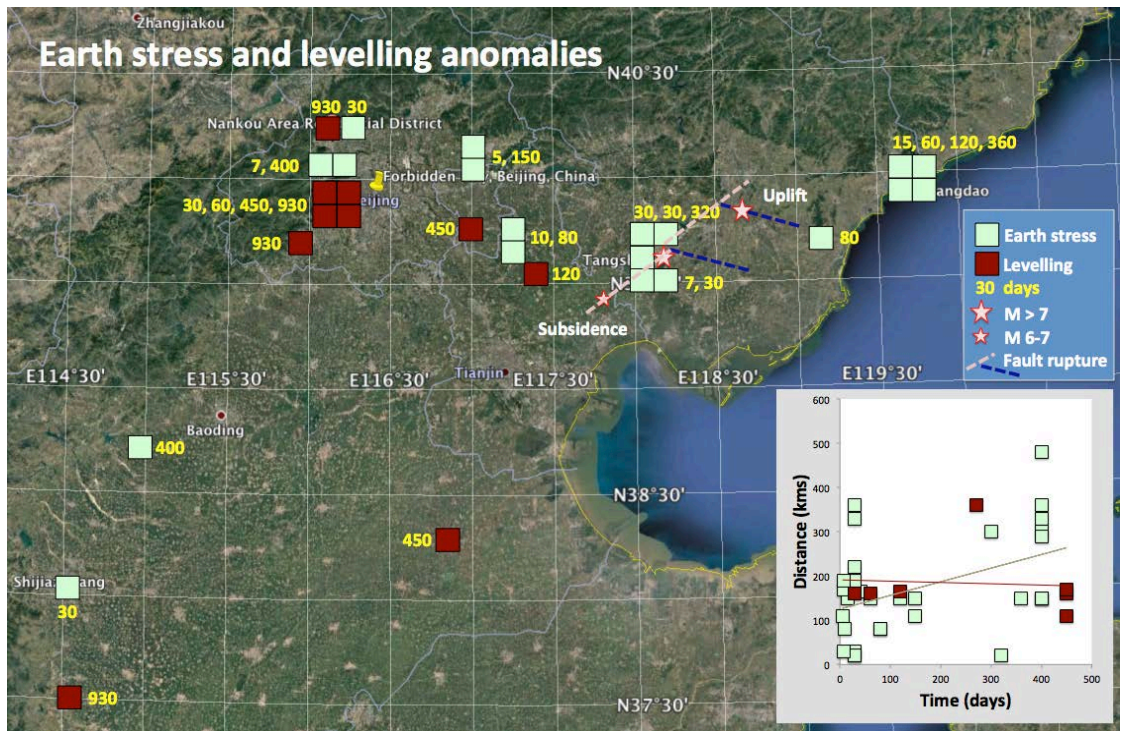

Figure 4. Map showing the distribution of Earth stress and levelling anomalies. See caption to Figure 3 for explanation of map annotations.

While there is a cluster of Earth stress anomalies on Tangshan, there is no clear focus on the city. If anything, there is a greater focus on Beijing. Once again, there is a tendency for these anomalies to focus on the basin margin and this may show that there was a change in stress between the basin and the basin margin. The inset graph shows that Earth stress anomalies drew closer to the epicentre as the time of the earthquake approached.

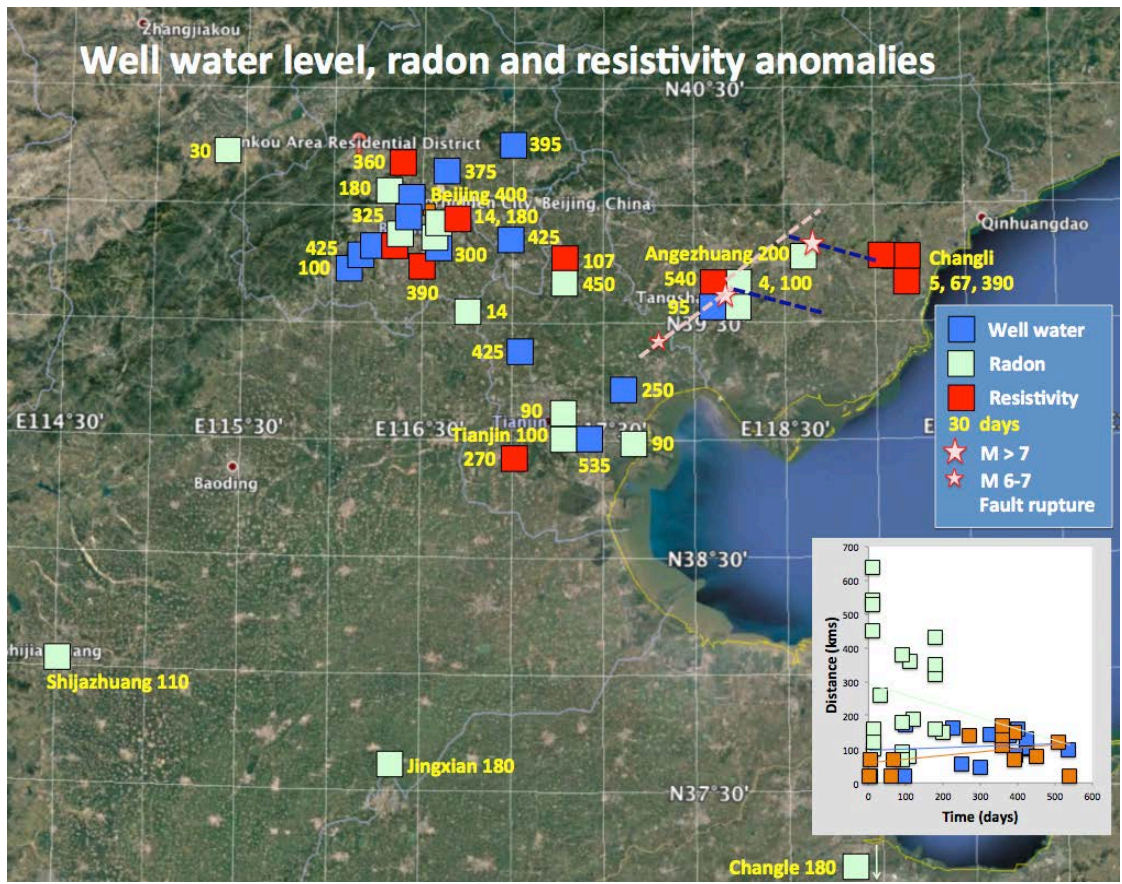

Figure 5. Map showing the distribution of well water level, well water radon and resistivity anomalies. See caption to Figure 3 for explanation of map annotations.

Well water level, well water radon and resistivity anomalies show a clear focus on the Beijing-Tianjin-Tangshan area but no clear focus on Tangshan itself (Figure 5). It is possible that weak lineaments in the distribution could be controlled by deep structure. The inset graph shows that near term radon anomalies appeared quite distant to the epicentre, >400 kms away. This contributed to confusion within the SSB.

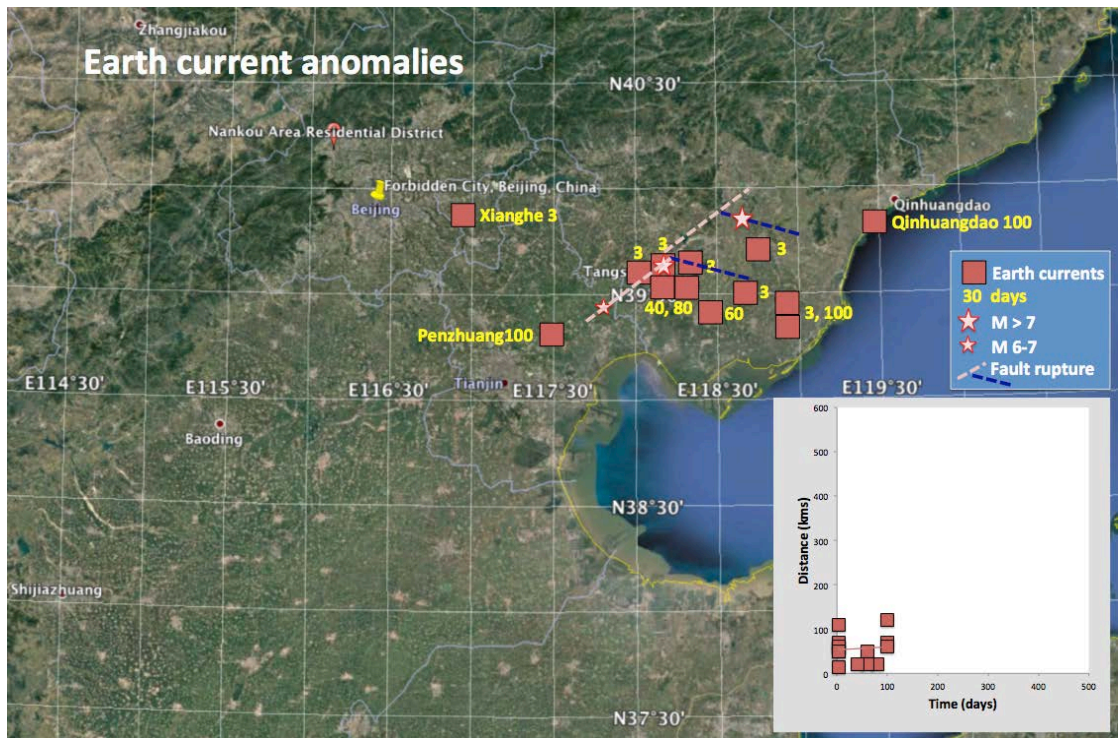

Figure 6. Map showing the distribution of Earth current anomalies. See caption to Figure 3 for explanation of map annotations.

Earth currents are the only anomaly class to be closely associated with the Tangshan earthquake in time and space (Figure 6). There are three spatial outliers at Xianghe, Qinhuangdao and Penzhuang. Earth current anomalies first appear 100 days before the earthquake, but at quite distant locations – Qinhuangdao, Baidaihe and Penzhuang, but then move towards the epicentre.

Earth current anomalies also show good spatial and temporal relationships with the Haicheng and Songpan earthquakes and could have been used to predict the Tangshan earthquake. Unfortunately, at the time, the Chinese seismologists were unaware of this relationship.

While the Earth currents provide good spatial resolution, questions would remain about timing. The objective of this paper is to find out if a finite time singularity model based on all the data may be used to improve temporal resolution.

### 3-Accelerated rates of precursor before the Haicheng and Tangshan earthquakes

### 3.1 Qualitative analysis

The Haicheng (1975/2/4) and Tangshan (1976/7/28) earthquakes were separated in time and space by 540 days and 450km. For each event, about 5 years of observation of anomalies are available leading up to the earthquake. For convenience, Table 1 summarises the list and number of anomalies by type in the 5 year period prior to the Tangshan earthquake, which are already discussed at the beginning of section 2. Such information is presently not available for Haicheng.

As visible from figures 7 and 8 and table 2, the frequency of anomalies clearly increases towards the earthquakes. More precisely, the null hypothesis that the 38 anomalies in the year prior to the Haicheng earthquake come from the same underlying frequency parameter as the 10 anomalies in the year before is rejected in favour of the alternative hypothesis that the frequency parameter underlying the 38 events is greater, with p=3x10$^{-5}$.

| Earth stress | Radon | Underground water | Resistivity | Earth currents | Gravity | Leveling | Geomagnetism |
|---|---|---|---|---|---|---|---|
| 31 | 29 | 21 | 20 | 15 | 14 | 11 | 11 |

Table 1. For Tangshan earthquake, the count of anomaly events by type, for those types with more than two anomaly events. Types with a single anomaly event include: 'energy of minor earthquake', 'hypocentral migration mechanism of minor earthquake', 'microseismicity', and 'oil flow' We have removed the Geodimeter (2 in total, one 9.8 yrs before and another 5.6 yrs before) and tiltmeter anomalies (1 in total, 2.6 yrs out). Keeping or removing these anomalies does not affect the statistical analysis later on, which uses data from more recent years.

Interestingly, Tangshan and Haicheng anomalies exhibit a similar acceleration, quantified by the share of their total anomalies in the final years: 16% and 17%, respectively, for one year before, and 63% and 65% for the year of the earthquake. Whether the total number of anomalies being quite different between the two cases is due to varying density of sensors or a lower actual rate of anomalies is unknown. Note that there is no clear indication of a spike in anomalies in the Tangshan series due to the Haicheng event, 540 days prior. However, e.g., resistivity and levelling anomalies do occur around that time.

| Years prior | 9 | 8 | 7 | 6 | 5 | 4 | 3 | 2 | 1 | 0 | Tot. |
|---|---|---|---|---|---|---|---|---|---|---|---|
| Tangshan | 0 | 0 | 0 | 3 | 1 | 5 | 8 | 15 | 25 | 102 | 159 |
| Haicheng | 0 | 0 | 0 | 5 | 0 | 0 | 2 | 3 | 10 | 38 | 58 |

Table 2. Number of anomalies in the ten 365 day periods up to the earthquake, going 10 years back, where the earthquakes is defined to occur at the end of year 0. The total number of anomalies is also given. The 7-9 are all zero and year 6 may be not very reliable with anomalies involving levelling and sea level. Since sea level was always changing, the corresponding anomalies are likely to be quite noisy.

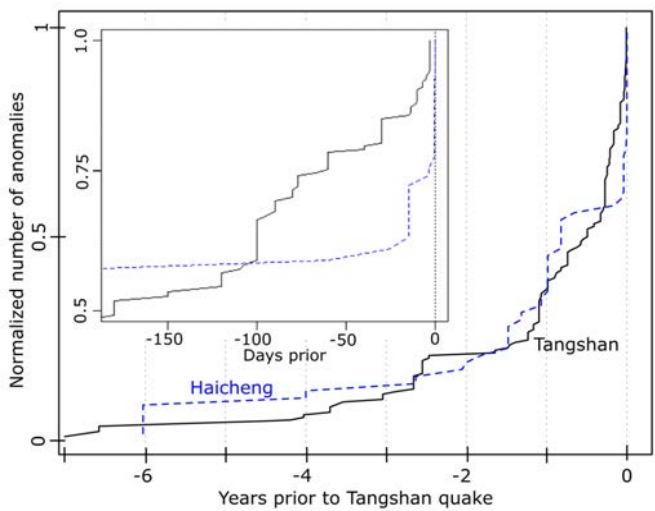

Figure 7. Normalized cumulative sum of anomalies in Tangshan (black) and Haicheng (blue dashed). Time goes from left to right. 0 corresponds to the occurrence time of the Tangshan (respectively Haicheng) earthquake. Inset: Six month time window version of these curves, which shows how Haicheng has a less clean LPPLS trajectory compared to Tangshan. See text for implications.

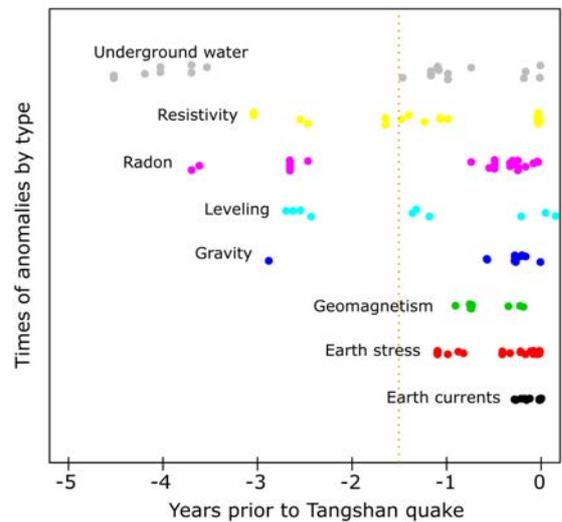

Figure 8. Replot of figure 2 of the time of anomalies leading up to the Tangshan earthquake, by type (excluding types with only a single anomaly event). Time goes from left to right as in figure 7, in units of years. The points of each anomaly type are offset vertically and scattered with a bit of noise for visibility. The time of the Haicheng event (540 days) is given by the vertical dashed line.

Next, it is possible that bursts or cyclical appearance of anomalies will exist in a baseline non-earthquake setting. Indeed, some of the anomalies occur in 2-3 apparent clusters (figure 8).

Seasonality in underground water and hence resistivity are expected. Earth current anomalies however only are reported within 100 days of the event. For all of the anomalies plotted in figure 8, with the exception of underground water, there is a significantly (p<0.05, using the R:poisson.test function) higher frequency of events in the year prior to the earthquake relative to the four years prior to that. This sample is inadequate to further characterize baseline behaviour, and data on additional earthquakes would be of great value.

## 3.2 Quantitative analysis

To enable further statistical analysis of the anomaly data, we (simply) combine all anomalies for each earthquake. As a flexible model of the acceleration of the rate of anomalies, we consider the LPPLS (Log-Periodic Power Law Singularity) (Anifrani et al., 1995; Saleur and Sornette, 1996; Sornette, 1998; Johansen and Sornette, 1998; Ouillon and Sornette, 2000), having event/point hazard rate,

$$h(t) \propto (t_c - t)^{-\alpha} \left(1 + \beta \cos(\omega \ln (t_c - t) + \phi)\right), \quad 0 < a < 1,$$  (1)

with a finite time singularity at $t_c$. This function, h(t), is the intensity of a Poisson process, having counting process N(t) — being the number of points up to and including time t. Its expected value is given by integrating the intensity,

$$E[N(t)] = \int_{-\infty}^{t} h(s)ds = \gamma + \delta \cdot (t_c - t)^{1-\alpha} \left(1 + \zeta \cos(\omega \ln (t_c - t) + \phi)\right), \, t \leq t_c.$$  (2)

Note that the condition a<1 prevents the finite time singularity in (2) from diverging. In other words, the rate diverges but the total number converges to a finite number at the singular time $t_c$.

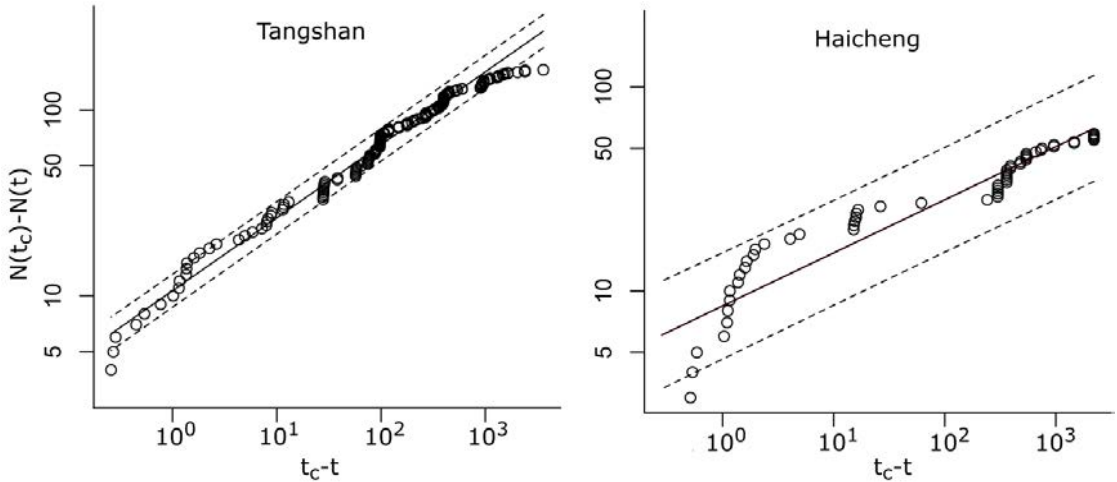

Figure 9. Diagnostic plot for log-linear relation eq. (3). For Tangshan (left plot), theoretical total number $N(t_c)$ of anomalies is taken as 165, where $t_c$ is taken as the day after the last anomaly. Eq. (3) with P(t)=1 (removing the log-periodic oscillations) is plotted with $\alpha = 0.75$. Time is given in days and flows from right to left. For Haicheng (right plot), $N(t_c)$ is taken as 60, $t_c$ is taken as the day after the last anomaly, and eq. (3) (removing the log-periodic oscillations) is plotted with $\alpha = 0.9$. The units of time are days, and double logarithmic scales are used to diagnose the log-linear relation in eq. (3). In both cases, theoretical 95% prediction intervals are plotted with dashed lines, however with residual iid assumptions clearly violated. The anomaly times were randomized with uniform noise on (-1,1) due to plausible data imprecision.

As a first diagnostic of LPPLS behavior, by specifying $N(t_c)$ and $t_c$, one can plot

$$N(t_c) - N(t) = c(t_c - t)^{1-\alpha} \ P(t),$$  (3)

where $P(t) := 1 + \zeta \cos(\omega \ln (t_c - t) + \phi)$ and look for a linear relationship in log-log scale of $N(t_c) - N(t)$ as a function of $t_c - t$, whose slope is equal to $1 - a$, up to log-periodic oscillations decorating the straight line. This is done in figure 9, where for Tangshan a value of $\alpha$ of about 0.75 is suggested, and for Haicheng about 0.9 (but with less data and a visually worse fit). The data and plot for Haicheng is still clean due to the long gap in points around 100 days out, and then the accumulation of points around 0-1 days out from $t_c$. To make it slightly cleaner, and also reasonable, we have randomized the date of the anomalies uniformly in (-1,+1) days.

We continue with Tangshan, having more data, making further statistical analysis feasible. We approach statistical estimation of the LPPLS (1) model, for sample $t_i$ i=1,…,n on window (l,r), as a probability density, with likelihood, $l(\theta) = \prod_{i=1}^{n} h(t_i)/\int_{l}^{r} h(s)ds$. This approach avoids: i) imposing a time scale of aggregation necessary for estimation via GLM regression; and ii) regression of function (2) to the observed counting process (cumulative number of anomalies), whose residual errors will be dependent, in violation of the regression assumptions. This method calibrating the data taken as a point process thus improves on existing methods previously used to qualify the presence of LPPLS and accelerated precursors (Anifrani et al., 1995; Johansen et al., 1996; 2000; Bowman et al., 1998; Guilhem et al. 2013).

The results of the MLE estimation of eq. (1) are summarized in Table 3 where time is measured in days. The result are relatively stable across the windows fitted – the first three fits going up to the time of the Tangshan earthquake but varying the length of anomaly history considered, and the last three varying the data to go up to 3, 5, and 10 days before the earthquake. The confidence intervals for the critical time all contain the true time (defined as $t_c = 0$), and are quite precise, with standard error of the parameter less than 1 day. In all cases, the log-periodicity component is highly significant (e.g., via t-test against $\beta$=0). The residual diagnostics indicate good fit of the trend (time scale transform test of Ogata (1988), checking for uniform residuals). A couple of the fits are visualized in figure 10.

| Sample size | Period | $\alpha$ | $\beta$ | $\omega$ | $\phi$ | $t_c$ |
|---|---|---|---|---|---|---|
| 126 | (-596,0) | 0.70 (.06) | 0.50 (.01) | 4.6 (.23) | 1.9 (1.2) | 0.5 (0.5) |
| 120 | (-447,0) | 0.61 (.06) | 0.51 (.11) | 4.9 (.41) | 3.9 (0.7) | 0.01 (0.01) |
| 100 | (-365,0) | 0.75 (.08) | 0.53 (.12) | 4.8 (.30) | 2.8 (1.4) | 0.9 (0.7) |
| 111 | (-596,-3) | 0.61 (.07) | 0.61 (.11) | 5.0 (.19) | 3.9 (.96) | -0.05 (.63) |
| 107 | (-596,-5) | 0.63 (.09) | 0.63 (.11) | 5.0 (.18) | 3.8 (.87) | -0.05 (.30) |
| 101 | (-596,-10) | 0.61 (.10) | 0.61 (.11) | 5.2 (.18) | 4.7 (.90) | 0.04 (.04) |

Table 3. Summary of fitted LPPLS hazard model (eq. 1) to Tangshan data, with estimated parameter value and standard error in parenthesis. Fits are given for six periods, including in the sample all anomalies within that given period, with 0 being the time of the earthquake.

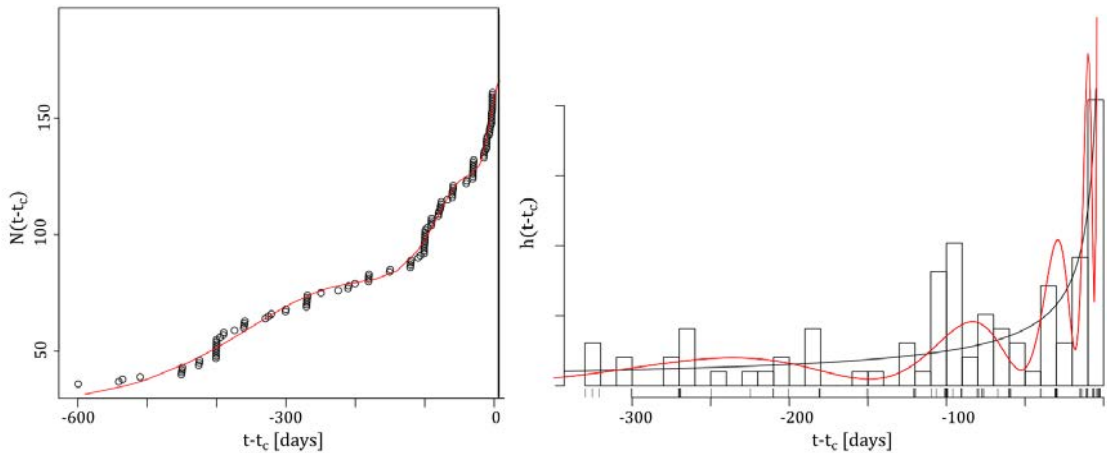

Figure 10. Cumulative number N(t-$t_c$) of anomalies as a function of days $t - t_c$ before the Tangshan earthquake. Time flows from left to right. Visualization of two LPPLS fits from Table 3. The left panel shows the fit from the first row of Table 3, in cumulative view (eq. 2), and the right the fit from the third row in hazard view or rate of anomalies (eq. 1).

Notably, the data available for the Haicheng earthquake gives much poorer results as shown in figure 11, which shows that the data is not well fitted by the LPPLS model. The fits are indeed very unstable due to the high clustering of points combined with large gaps between clusters, as well as the small sample size. The hazard function would have to be allowed to go negative to allow better fitting. This may come as a surprise, given the fact that the Haicheng earthquake was successfully predicted by the SSB and is the only official prediction that led to a large-scale evacuation in human Earthquake history, while the Tangshan earthquake was not officially predicted. Post-mortem analyses have shown, however, that the Haicheng forecast was based mainly upon the seismic foreshock sequence (Raleigh et al., 1977), which is not taken into account in the present work.

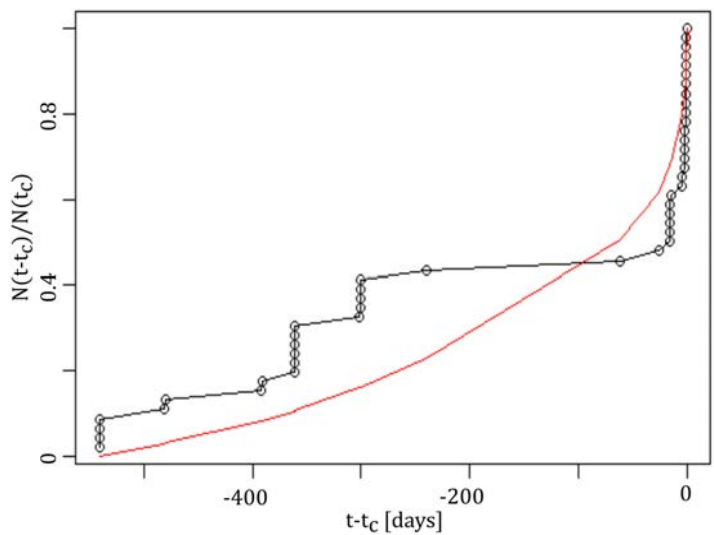

Figure 11. Haicheng best LPPLS fit, cumulative view: total number of anomalies up to time t shown on the x-axis, where time (in days) flows from left to right with 0 as the time of occurrence of the Tangshan earthquake. Fit is to be taken with caution. Uniformity of residuals test rejects the LPPLS model with p=0.0047. Estimated parameters are $\alpha = 0.7, \beta = 0.2, \omega = 5, \phi = 3, t_c = 0.01$. Standard errors were not numerically possible.

### 3.3 Forecasting (ex-post) the Tangshan earthquake and testing its predictability

To more fully examine the evolution of the predicted critical time when approaching the Tangshan earthquake, we fit the LPPLS hazard function (eq. 1) on a growing window (l=-596, r) with r= -50,-49,…,0, (unit of days) and again with l= -365 (unit of days), where the earthquake time is always defined to be at time 0. The 90% profile likelihood confidence interval for the predicted critical time is summarized in figure 12 for both of these cases, which does a reasonable job of bracketing the true critical time. The prediction is highly uncertain more than 30 days before the event, however, from about 20 days before, the prediction becomes increasingly precise as the earthquake time is approached – with a slight shift towards a too early estimate of the earthquake time (about 5-10 days). With the exception of a few unstable fits, the prediction of an imminent event becomes clear in the week prior to the event.

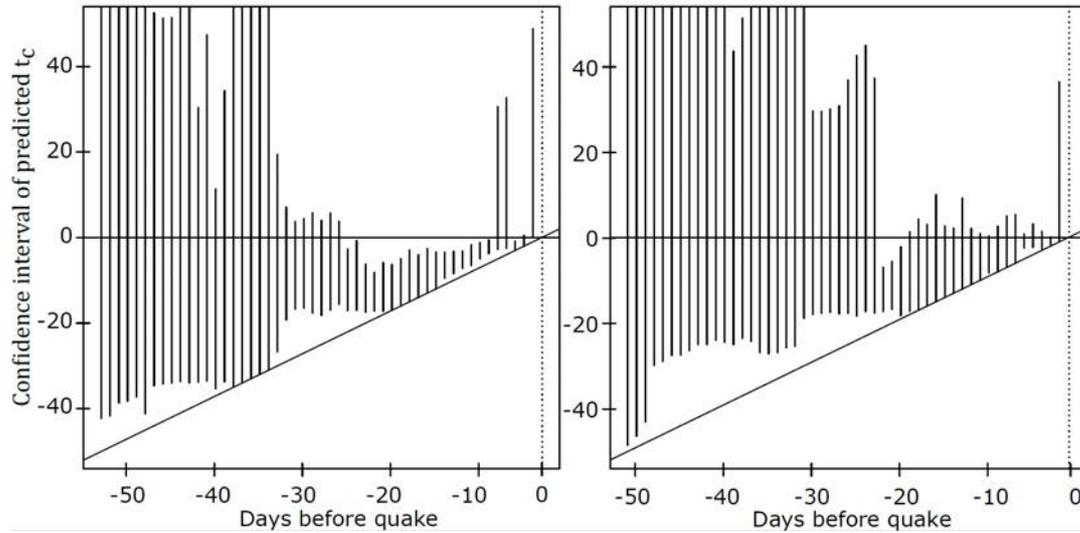

Figure 12. 90% confidence interval of predicted time $t_c$ (earthquake occurrence time) in days vs. days before earthquake. Along the x-axis and y-axis, 0 corresponds to the time of occurrence of the Tangshan earthquake. The x-axis gives the "present" time at which the mock prediction is made. For instance, at 10 days before the earthquake, the predicted time of the earthquake is between -1 and -3 days, i.e. the method predicts the event between 1 to 3 days before the earthquake actually occurred. The diagonal line gives a lower bound on the estimate which is where the estimate of the critical time is the end of the fitted data window (i.e. the "present" time). Note that when the confidence interval lines go above 0, the critical time is overestimated and vice versa. The left plot fixes the data starting time at l= -596 days prior to the earthquake, while the right plot takes it at l= -365. Dramatic changes in the confidence interval are possibly due to the periodicity of the function, which can inject bi-modality into the likelihood over the critical time parameter.

We now perform a simulation study to test the feasibility of an early warning system if the anomalies really do follow an LPPLS trajectory. In particular, assuming parameters consistent with the observed data (taking parameters from row 1 of table 3), we indicate the predictability of the earthquake time with a simple prediction experiment: for a given realization of anomalies (data), we estimate expression (1) each day in the 40 days up to $t_c$, fitting on the growing sample of observed anomalies $\{t_1, ..., t_{N(i)}\}$, for days i=86,...,126. We declare an alarm at the first individual fit in the sequence of 40 where the predicted critical time (estimated parameter $t_c$) is sufficiently near and precise. In detail, we declare an alarm if the nearest edge of the 95% confidence interval of $t_c$ falls within 7 days of the most recent date $t_i$ within the sample (i.e., 'today' in a real-time prediction context), and its width is less than 14 days. We can also count the total number of the 40 fits that satisfy the alarm condition. Simulations indicate that, in ~93% of cases, alarm will be raised for an earthquake within the 10 days before the actual time. Also, in ~5% of cases, an alarm will be prematurely raised for earthquakes falling in the interval from 40 to 20 days prior to the actual time. A rate of false alarm is not quantified here, given lack of observation of anomaly data under non-earthquake conditions. However, under a homogeneous Poisson assumption for anomaly times for non-earthquake conditions, the observed super-exponential accumulation of points is exceedingly unlikely: for instance, the Kolmogorov-Smirnov test that the point process of anomalies is uniform rejects this null with p<10^-16.

## 4. Conclusion

Not widely known, the Chinese, under the stewardship of Enlai Zhou, in the period 1966 to 1976, developed an earthquake prediction methodology in one of the world's biggest science and social science projects ever undertaken. The program climaxed with the successful prediction of the M7.6 Haicheng earthquake in February 1975 that saved many thousands of lives. Eighteen months later, the industrial city of Tangshan, 180 kms east of Beijing, was flattened without warning, by an M7.8 earthquake. Somewhere between 240,000 and 650,000 people lost their lives. The jubilation that followed Haicheng success turned to despair. Zedong Mao died in September 1976. China set a new course and the earthquake prediction program was prematurely dismantled and all but forgotten. The recipe for success lay smoldering in hundreds of technical documents and reports.

Here, we have revisited the compiled set of precursory data that were reported to be available in real time before the Haicheng and Tangshan earthquakes. Rather than analyzing each anomaly one by one (geodesy, levelling, geomagnetism, soil resistivity, Earth currents, gravity, Earth stress, radon, well water level), we have proposed a coarse-graining approach consisting in simply counting the total number of reported anomalies as a function of time. We have demonstrated a strong evidence for the existence of an acceleration of anomalies leading up to the major Haicheng and Tangshan earthquakes. In particular for the Tangshan earthquake, the frequency of occurrence of anomalies is found to be well described by the LPPLS (log-periodic power law singularity) model.

Based on a mock real-time prediction experiment, and simulation study, we have provided an indication of potential for an early warning system based on this methodology of monitoring accelerated rates of anomalies.

Further data from these and additional earthquakes would be of great value to better characterize both baseline and earthquake anomaly behavior. Indeed, the set of relevant anomalies, and the statistics of anomalies near an earthquake may vary considerably by region and case, notably due to differing geology and tectonic settings.

It is likely that the acceleration that we report here was somehow understood implicitly by the Chinese experts gathering and trying to make sense of the accumulating data sets. While such "gut feelings" may explain the human drama of how scientists and decision makers were discussing and quarrelling over the growing feeling of incipient risks, the experts apparently never made the step to quantify this impression of an imminent danger in the simple and transparent way that we have proposed here. In fact, the data from hundreds of monitoring stations was never collated into a central database and the Chinese at the time depended upon amateur staff manning individual stations to draw and report conclusions based upon that single station's results. In the spirit of optimally displaying technical information (Tufte, 2001), we propose that such simple robust metric as proposed in the present paper may often bring substantial evidence to a developing trend and the impending dangers. Perhaps, this could have convinced the decision makers to act. We see here a parallel to the story of the 1986 Challenger disaster, for which the engineers of the company Thiokol were unable to convey with sufficient clarity to their managers and those of NASA why they were opposed to the take-off on that fatal day. Tufte (2001) has shown that an efficient visual display (Figure 13) of the trend of the damage of the O-ring of the boosters on the side of the main rocket could have changed the decisions, which were at that time based on arcane reports and tables full of incomprehensible numbers (for the managers and decision makers).

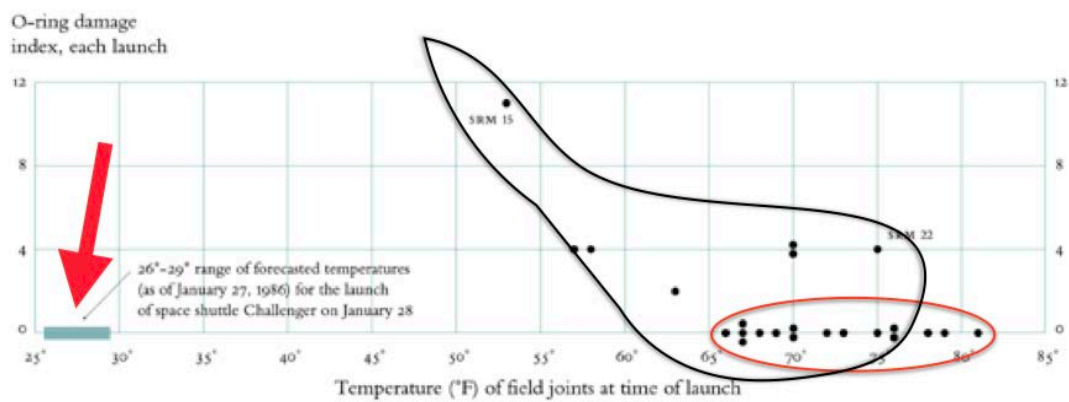

Figure 13. Measure of damage of the O-ring rubber on the booster rockets on the side of the main rocket carrying the space shuttle as a function of temperature (in Fahrenheit) at take-off of previous flights. Each black dot is a measure taken for a different previous flight. The red arrow indicates the range of temperature at take-off on January 28, 1986, when the Space Shuttle Challenger broke apart 73 seconds into its flight, killing all seven crew members aboard. Note that this was below 32°F, the temperature at which water freezes. The red ellipsis emphasizes the importance of reporting data even when no damage is observed ("the dog that did not bark"),

which occurred for flights taking off at relatively warm temperatures on the ground. The rocket-like envelop represents the noisy trend of the growing number of previous flights that have shown significant damage of the critical O-ring structure as the temperature at take-off was lower. Notice the large gap between the temperature at take-off of all previous flights and that of the fateful January 28, 1986. Given this data evidence, would you have taken the decision to authorize the launch? Detailed post-mortem analyses confirmed the causal role of the failure of O-rings in the Challenger disaster (adapted from Tufte, 2001).

**References:**

Anifrani, J.-C., C. Le Floc'h, D. Sornette and B. Souillard, 1995. Universal Log-periodic correction to renormalization group scaling for rupture stress prediction from acoustic emissions, J.Phys.I France 5 (6), 631-638.

Bowman D.D., Ouillon G., Sammis C.G., Sornette A., Sornette D., 1998. An observational test of the critical earthquake concept, J. geophys. Res. 103 B10, 24349-24372.

Guilhem, A., R. Bürgmann, A.M. Freed and S. Tabrez Ali, 2013. Testing the accelerating moment release (AMR) hypothesis in areas of high stress, Geophysical Journal International 195 (2), 785-798.

Gluzman, S. and D. Sornette, 2002. Classification of Possible Finite-Time Singularities by Functional Renormalization, Physical Review E 6601 (1), art. no. 016134, PT2:U315-U328.

Goldenfeld, N., 1992. Lectures On Phase Transitions And The Renormalization Group (Frontiers in Physics), CRC Press, Boca Raton.

Ide, K. and D. Sornette, 2002, Oscillatory Finite-Time Singularities in Finance, Population and Rupture, Physica A: Statistical Mechanics and its Applications 307 (1-2), 63-106.

Jiang, H. and Chen, F. (editors), 1982. The 1976 Tangshan Earthquake. Seismological Press Beijing, Second Printing. (in Chinese).

Johansen, A. and D. Sornette, 1998. Evidence of discrete scale invariance by canonical averaging, Int. J. Mod. Phys. C 9, 433-447.

A. Johansen, H. Saleur and D. Sornette, 2000. New Evidence of Earthquake Precursory Phenomena in the 17 Jan. 1995 Kobe Earthquake, Japan, Eur. Phys. J. B 15, 551-555.

Johansen, A., D. Sornette, H. Wakita, U. Tsunogai, W.I. Newman and H. Saleur, 1996. Discrete scaling in earthquake precursory phenomena : evidence in the Kobe earthquake, Japan, J.Phys.I France 6, 1391-1402.

Mei, S., 1986. The precursory complexity and regularity of the Tangshan earthquake. J. Phys. Earth, 34, 193-212.

Mearns, E. and D. Sornette, 2020. To be published

Mignan A., 2011. Retrospective on the Accelerating Seismic Release (ASR) hypothesis: controversy and new horizons, Tectonophysics 505, 1-16.

Ogata, Y., 1988. Statistical models for earthquake occurrences and residual analysis for point processes, Journal of the American Statistical association 83 (401), 9-27.

Ouillon, G. and D. Sornette, 2000. The Concept of 'critical earthquakes' applied to mine rockbursts with time-to-failure analysis, Geophys. J. Int. 143, 454-468.

Qian, F.-Y., Y.-L. Zhao and J. Lu, 1997. Georesistivity precursors to the Tangshan earthquake of 1976, Annali di Geofisica 40 (2), 251-260.

Raleigh, B., G. Bennett, H. Craig, T. Hanks, P. Molnar, A. Nur, J. Savage, C. Scholz, R., Turner and F. Wu, 1977. Haicheng earthquake study delegation, Prediction of the Haicheng Earthquake, EOS Trans. Amer. Geophys. Union 58, 236-272.

Saleur, H. and D. Sornette, 1996. Complex exponents and log-periodic corrections in frustrated systems, J.Phys.I France 6, n3, 327-355.

Sichuan Province Earthquake Office, 1979. The 1976 Songpan Earthquake (in Chinese, authors unkown).

Sornette, D., 1998. Discrete scale invariance and complex dimensions, Physics Reports 297 (5), 239-270 (extended version at http://xxx.lanl.gov/abs/cond-mat/9707012).

Sornette, D., 2002. Predictability of catastrophic events: material rupture, earthquakes, turbulence, financial crashes and human birth, Proceedings of the National Academy of Sciences USA 99 (Suppl. 1), 2522-2529.

Sornette, D., 2004. Critical Phenomena in Natural Sciences (Chaos, Fractals, Self-organization and Disorder: Concepts and Tools), 2nd ed., pp.528 (Springer Series in Synergetics, Heidelberg).

Sornette, D. and C.G. Sammis, 1995. Complex critical exponents from renormalization group theory of earthquakes : Implications for earthquake predictions", J.Phys.I France 5, 607-619.

Tufte, E., 2001, The Visual Display of Quantitative Information, Graphics Press; 2nd edition edition.

Zheng, Jian-zhong, 1981. Precursors to the Haicheng and Tangshan earthquakes, Earthquake Part 2, 34 (1), 43-59 (in Chinese).

Zhu, F. M. and Wu, G., editors, 1982. *The Haicheng Earthquake of* 1975. Seismological Press, Beijing, 220 pp. (in Chinese)